\documentclass{sig-alternate-10pt}

\usepackage{graphicx}
\usepackage[hyphens]{url}
\usepackage{multicol}
\usepackage{array}
\usepackage{booktabs}
\usepackage{multirow}
\usepackage[table]{xcolor}
\usepackage{tikz}
\usepackage{microtype}

\begin{document}

\title{\Large \bf TLS Proxies: Friend or Foe?}

\numberofauthors{1}

\author{                                                                        \alignauthor
Mark O'Neill, Scott Ruoti, Kent Seamons, Daniel Zappala\\
\affaddr{Brigham Young University}\\
\affaddr{Department of Computer Science}\\
\affaddr{Provo, UT 84602}\\
\email{\normalsize mto@byu.edu, ruoti@isrl.byu.edu, seamons@cs.byu.edu, zappala@cs.byu.edu}
}

\maketitle

\begin{abstract}

The use of TLS proxies to intercept encrypted traffic is controversial
since the same mechanism can be used for both benevolent purposes,
such as protecting against malware, and for malicious purposes, such
as identity theft or warrantless government surveillance. To
understand the prevalence and uses of these proxies, we build a TLS
proxy measurement tool and deploy it via Google AdWords campaigns.
We generate 15.2 million certificate tests across two large-scale measurement studies.  We find that 1 in 250 TLS
connections are TLS-proxied. The majority of these proxies appear to be
benevolent, however we identify over 3,600 cases where eight malware
products are using this technology nefariously.  We also find numerous
instances of negligent, duplicitous, and suspicious behavior, some of which degrade
security for users without their knowledge.  Distinguishing these
types of practices is challenging in practice, indicating a need for
transparency and user awareness.

\end{abstract}

\section {Introduction}
\label{sec:introduction}

Secure communication on the Internet is based primarily on digital
certificates signed by certificate authorities and intermediate
authorities. This validation system is currently being compromised by
the use of TLS proxies, which can act as a man-in-the-middle (MitM)
for TLS connections (e.g., \cite{bluecoat, paloalto, adlocksmith,
  bluecoatproxysg, symantecwebgateway}).  A TLS proxy can issue a {\em
  substitute certificate} for any site the user visits, so that the
user establishes an encrypted connection to the proxy rather than the
desired web site.  The proxy can then decrypt and monitor or modify
all user traffic, before passing it along via a second encrypted
channel to the desired web site. TLS proxies are used for a variety of
legitimate purposes, such as blocking malware, but can also be used by
malicious entities to compromise the privacy or security of end users.
Isolated attacks had been observed in the wild, notably in Iran
\cite{iranattack} and Syria \cite{syriaattack}.  The most dangerous
aspect of TLS proxies is that the user is entirely unaware that
encrypted traffic is being intercepted by an organization or
attacker. The use of TLS proxies is controversial because browser
software still shows a lock icon during such sessions, misleading
users and compromising the end-to-end security promises made by TLS.

Detecting the presence and prevalence of TLS proxies is a challenging
measurement problem. To detect a proxy, we must obtain the certificate
a client, such as a web browser, actually obtains, then compare this
with the valid certificate presented by the server the client is
contacting. A mismatch indicates that some kind of proxy, either
benevolent or malicious, is intercepting the client's traffic to that
particular server. To determine the prevalence of TLS proxies, we must
repeat this measurement on as many client systems as possible. A
variety of studies have examined the certificate ecosystem by scanning
secure servers from a single point of view
\cite{holz2011ssl,eckersley2011decentralized,durumeric2013analysis} or
using passive monitors from several vantage points
\cite{eckersley2011decentralized,amann2013acsac}. However, detecting
proxies necessitates measurements at the client, and less work has
been done in this space.


Two recent works have found some evidence for TLS proxies by measuring
certificates received by clients. Huang et al. measure the prevalence
of TLS proxies that intercept traffic from clients connecting to
Facebook \cite{huang2014analyzing}, finding that 1 in 500 TLS
connections are proxied, mostly by corporate Internet filters and
personal antivirus software. In addition, a small number of
connections were found to be intercepted by malware. Because this
study uses Flash to detect a certificate mismatch, it does not detect
proxies affecting most mobile devices. The Netalyzer project measured
certificates received by Android apps, assessing 15,000 sessions and
identifying just one case of a TLS proxy running in an analytics app
\cite{vallina2014tangled}. Though this is a very low rate of
prevalence (30 times less than Huang's study), the app was found to
whitelist several sites, including Facebook. This indicates that
measurements of proxies should examine low-profile sites that are
unlikely to be whitelisted.

In this paper, we report on our measurements of the prevalence of
proxies using a Flash app deployed with a Google AdWords
campaign. Like Huang, our measurements use Flash to detect a
certificate mismatch without any user interaction. However, we deploy
our tool using a Google AdWords campaign, which affords a number of
advantages. First, we are able to actively measure clients, based on
how much money we spend on the advertisement, enabling us to collect
as many as 12 million measurements in one week by spending \$750 per
day. Second, we are able to target our measurements at any country, so
that we can measure proxy prevalence in distinct areas of the
world. Third, we are able to measure any site that has a permissive
Flash socket policy file. Together, these characteristics give us a
broader view of TLS proxies on the Internet, including those
that may have whitelisted Facebook.

In this paper, we report on a two-part measurement study of TLS
proxies using a Google AdWords campaign.  The first part of the study
measures proxies broadly, wherever Google places our advertisement,
comprising 2.9 million certificate tests, with proxied users in 142
countries.  The second part of the study specifically targets users in
five countries (China, Ukraine, Russia, Egypt, and Pakistan) in
addition to the world at large.  This covers 12.3 million certificate
tests, finding proxied users in 147 countries.  The first part
measures proxies intercepting traffic to a new server on our campus,
which is highly unlikely to be whitelisted by any proxy.  The second
part measures proxies intercepting traffic to seventeen sites on the
Alexa top million in addition to our own server.  We are able to scan
these seventeen sites because they have permissive Flash socket policy
files; scans of the entire Alexa top million are not possible without
installing new software on client machines.

Our basic findings are as follows:

\begin{itemize}

\item The first part of our study found 11,764 proxied connections out
  of 2.9 million total measurements (0.41\% or approximately 1/250 of
  all connections) spanning 142 countries.  This rate is
  double that reported by Huang.  We found that most
  substitute certificates claim to be from benevolent TLS proxies,
  with 70.87\% claiming to be generated by a firewall software and
  12.66\% claiming to be generated by a corporate network.

\item The second part of our study, which queried multiple secure
  hosts, found 50,761 proxied connections out of 12.3 million total
  measurements (again, 0.41\% of all connections) spanning 147
  countries. It is surprising that the overall prevalence is identical in
  both studies, which seems to indicate that none of the sites we tested
  was whitelisted by proxies.

\item Our second study targeted specific countries with the Google
  AdWords placement.  We find that proxy rates vary significantly with
  respect to the origin country of the user. China has an exceptionally
  low rate of TLS proxies whereas the United States and other western
  nations tend to have much higher rates of TLS proxies. Targeted
  countries also have a greater rate of unclassifiable TLS proxies
  that disclose little to no information about their nature.

\item In both studies we found numerous instances of negligent and
  malicious behavior. Our analysis of one parental filter finds that
  it masks forged certificates, allowing an attacker
  to easily perform a MitM attack against the firewall's users.  In
  addition, we found eight malware products affecting over 3,600
  connections that install a new root certificate and act as a TLS
  proxy to dynamically insert advertisements on secure sites.  We also
  found evidence that spammers are using TLS proxies in their products
  and that botnets may be using this technique. We found numerous
  other suspicious circumstances in substitute certificates, such as a
  null Issuer Organization, falsified certificate authority
  signatures, and downgraded public key sizes.
 
\end{itemize}

\section{Background}

\begin{figure}
\centering
\includegraphics[width=\columnwidth]{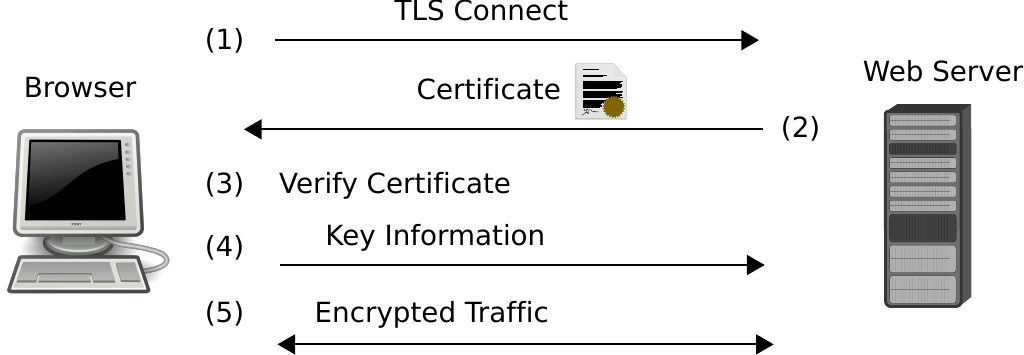}
\caption{High-level view of secure session establishment via TLS}
\label{fig:certificate}
\end{figure}

To validate the identity of a website such as Amazon, the web browser
relies on certificate authorities (CAs), which digitally sign
certificates vouching for the identity of the web server. 
Figure~\ref{fig:certificate} provides a high-level overview of a 
TLS handshake where the browser initiates a TLS connection
with the server, and retrieves the server's certificate. The browser
verifies this certificate and, if it is valid, sends random 
data securely to the server that is used to seed the generation of 
the necessary encryption keys.
The browser and the server can now exchange encrypted traffic.

\begin{figure}
\centering
\includegraphics[width=\columnwidth]{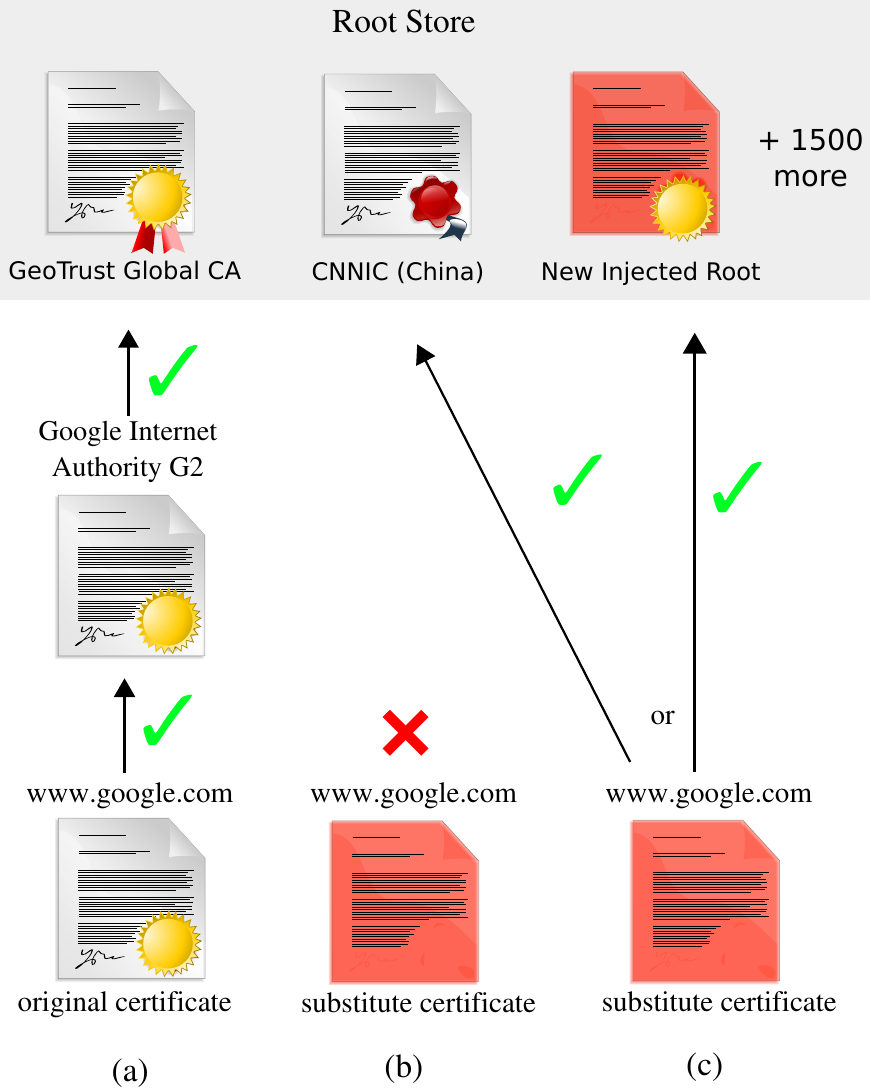}
\caption{Certificate validation. Part (a) shows expected behavior when the www.google.com webserver provides its certificate to the browser for validation. The certificate's signature is validation against an intermediate certificate, Google Internet Authority G2, which is in turn signed by a certificate in the trusted root store.  Part (b) shows a substitute certificate invalidated due to lack of any signatures leading back to a root certificate. Part (c) shows a validated substitute certificate issued by a rogue or compromised certificate authority (left arrow) and a validated substitute certificate signed by an authority who injected a new trusted certificate into the root store.}
\label{fig:certificatevalidation}
\end{figure}

Web browsers authenticate a site by validating a chain of digital signatures from the site's certificate back to one of a set of trusted root certificates. These certificates comprise the ``root store'' and are
typically bundled with the operating system or browser. For example,
the certificate for www.google.com is signed by the Google Internet
Authority G2, an intermediate certificate authority run by
Google. This certificate is in turn signed by GeoTrust Global CA, a
certificate authority whose certificate is located in the root store
of the browser or operating system.
This is exemplified in Figure~\ref{fig:certificatevalidation}(a). The legitimate certificate for google.com is accepted because it has a valid chain of signatures leading back to a trusted certificate in the root store.
Figure~\ref{fig:certificatevalidation}(b) shows a case where a substitute certificate's signature cannot be traced back to a root store certificate and is rejected.
This is expected behavior for substitute certificates.

\begin{figure}
\centering
\includegraphics[width=\columnwidth]{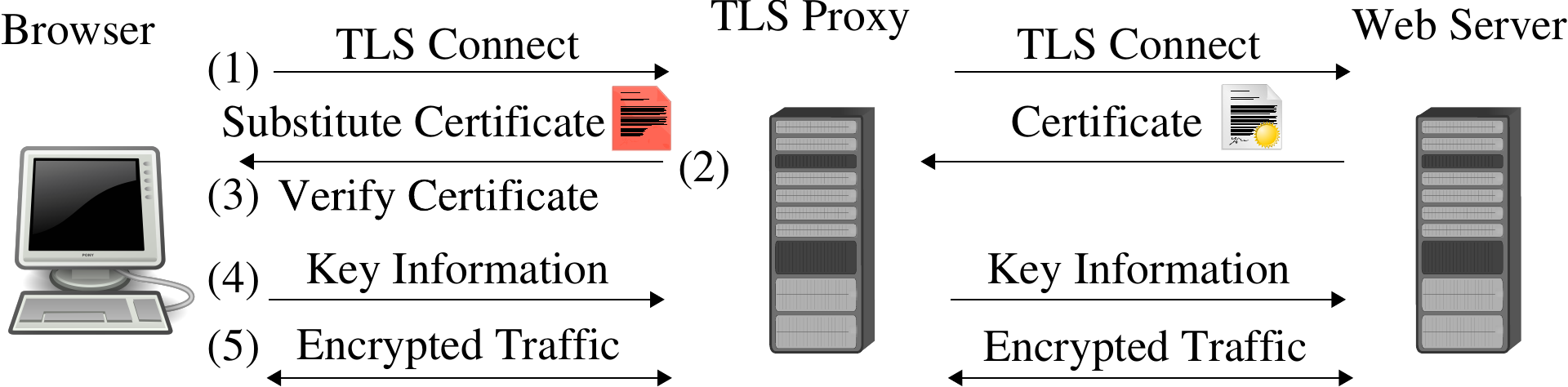}
\caption{``Secure'' session establishment with involving a TLS proxy}
\label{fig:proxy}
\end{figure}

This system can be attacked by a TLS proxy inserting itself as a
man-in-the-middle between the browser and the web server. As shown in
Figure~\ref{fig:proxy}, when the browser tries to open a secure
connection to the web server, this connection is instead intercepted
by the proxy. The proxy also provides a falsified, substitute
certificate to the browser, so that it can impersonate the
original website. For this to work, the proxy must somehow control a substitute certificate for the original website that validates against the root store of the user. This can be accomplished in a variety of ways, both benign and malicious.  

Figure~\ref{fig:certificatevalidation}(c) illustrates two ways the operator of a proxy can provide valid
certificates that do not elicit a browser warning. One way (indicated by the vertical arrow) is to directly insert a certificate into the trusted root store. For example, an
organization can use enterprise software to supply its own
certificates for the root store of all computers in the organization.
Alternatively, the IT department of an organization can create a
software image using new root certificates and require that all
computers in the organization use this image. This is generally done
for benevolent reasons, such as blocking malware and viruses,
providing intrusion detection, or protecting intellectual
property. Likewise, personal firewall software often installs a new
root certificate so that it can offer users the feature of scanning
their encrypted traffic for malware. Finally, in the case of mobile
devices, a manufacturer may add certificates to the root store,
ostensibly for speeding up web browsing.  Nokia came under fire
recently for doing this \cite{nokia}.

Other ways of providing seemingly-valid certificates are more nefarious. A malicious entity can also install new certificates into an operating system's root store. For example, malware typically
has permission to add new certificates when it is installed
inadvertently by the user. Sometimes this may be as simple as asking
the user for install privileges, and many users are trained to click
"OK" whenever the operating system asks them for this.  It is even possible for a malicious entity to act as a proxy without adding any
new certificates to a victim's root store. 
This is shown by the left-leaning arrow from the certificate in Figure~\ref{fig:certificatevalidation}(c). A rogue certificate
authority can issue any certificate it wants, since all root certificates are allowed to sign for any domain. There have also been numerous reported cases of compromised and negligent certificate authorities that allow attackers to issue fraudulent certificates \cite{durumeric2013analysis}.
In addition, some governments, such as China and the United States, control their own root certificate
authorities, and evidence suggests that a government may coerce
authorities into granting them certificates for domains they do not
own \cite{marlinspike2011ssl, eckersley2011decentralized}.

In summary, a TLS proxy can be operated by both malicious and benign parties including parental control software, corporate firewalls, malware, hackers, and government surveillance operatives.

\section{Measurement Tool}

We have developed a tool to measure the prevalence of TLS
proxies using existing, widely-deployed technologies.  The tool runs silently from the
perspective of the user; no user action is required, either to
install any software or to interact with the tool.  This is a
significant advantage as compared to other work that requires
client-side software installation \cite{wendlandt2008perspectives,
  marlinspike2011ssl, alicherry2009doublecheck,holz2012x,
  amann2012extracting, amann2012revisiting}.  

\subsection{Design}

To meet our objective of using existing browser technologies without requiring further client installation, we take advantage of the widespread deployment and transparency afforded by the Adobe Flash runtime.
By hosting a Flash application on a web page the server can upload it to a visiting client, which runs it without any user interaction.
Our tool works by sending a \texttt{ClientHello} message to a TLS-enabled server and recording the \texttt{ServerHello} and \texttt{Certificate} messages received in response. The retrieved certificate(s) is then forwarded to the web server for verification.
This process is handled in three steps, illustrated in Figure~\ref{fig:flashtool}:

\begin{figure}
\centering
\includegraphics[width=\columnwidth]{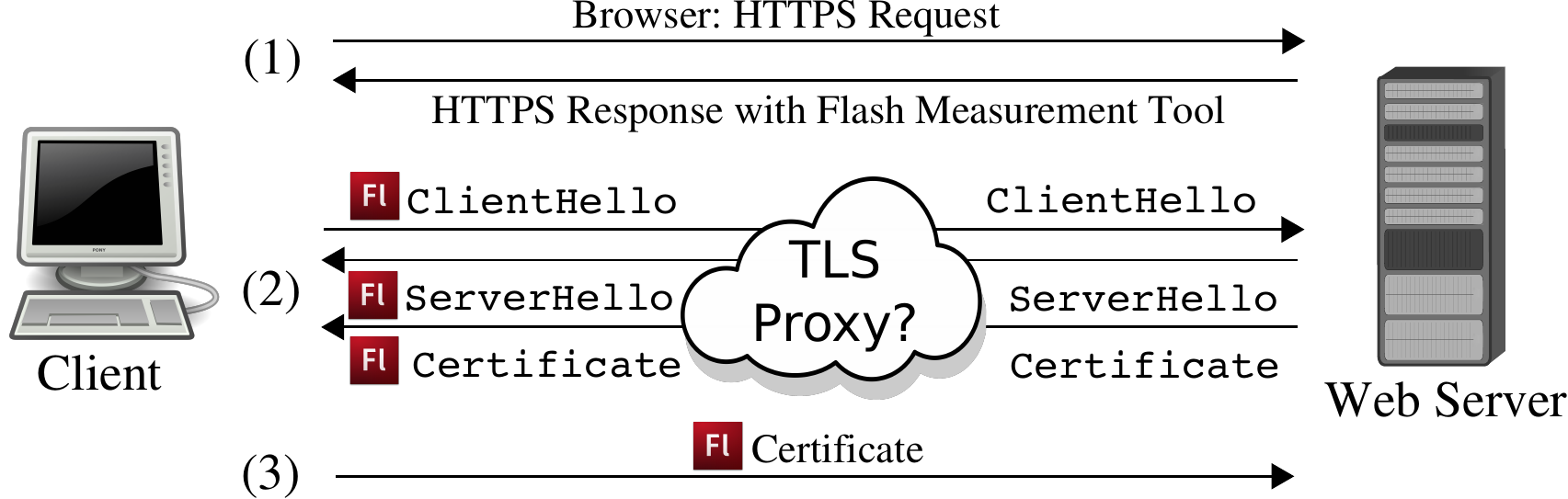}
\caption{TLS Proxy Measurement}
\label{fig:flashtool}
\end{figure}

\begin{enumerate}

\item {\bf Retrieve measurement tool.} The client browser connects to the web server where the Flash application is hosted.
The application need not be visible to the user and can merely be embedded in the background of an otherwise normal web page.
The web page data, along with the embedded Flash application, is then downloaded by the client.

\item {\bf Record certificate.} The Flash tool is run automatically by the browser. The tool sends a \texttt{ClientHello} message back to the server to initiate a TLS handshake.
The tool then records the \texttt{ServerHello} and \texttt{Certificate} messages received in response and terminates the handshake.

\item {\bf Report results.} The tool reports these results back to the server using an HTTP POST request. The server then compares the certificate received with the original it sent.  A mismatch indicates the presence of a TLS proxy.

\end{enumerate}

To deploy our tool a website administrator need only do the following:

\begin{enumerate}

\item {\bf Host the Flash application on the web server } and invoke it from the desired pages. The tool can be deployed transparently within existing web pages with no visible changes.

\item {\bf Host a simple ``socket policy file'' on the server.} For security reasons the Flash runtime requires that applications attempting to establish a TCP connection with a remote host first obtain permission from that host via a simple policy file. The request for this file is sent automatically by the Flash runtime. The software to host such a file is extremely simple and easily deployed. This particular security feature of Flash prohibits the tool from testing client connections to arbitrary hosts; all hosts tested must first grant permission through their respective socket policy files. We serve our socket policy file on the same port used by our web server (80). This reduces the effect of captive portals, which often block traffic targeting ports other than those used by HTTP and HTTPS (e.g., airport public access WiFi). Our socket policy server implementation is provided on our website.

\end{enumerate}

\subsection{Implementation}

To implement our tool it was necessary to retrieve the certificate used during a TLS handshake.
It would have been preferable to use JavaScript or HTML5 to retrieve the certificate used as part of a current TLS connection, but unfortunately there is no API available for this. Firefox allows a plugin to request the certificate, but plugins require manual client installation.
This left us with the alternative of establishing a plain TCP connection with the target server and then initiating a TLS handshake. Unfortunately, the ability to use a plain TCP connection rules out the use of HTML5  \texttt{WebSockets}.

Due to these constraints, we opted to use the Adobe Flash platform.
Beginning with version 11.0 of the Flash runtime, Adobe made available a \texttt{SecureSockets} API that allows developers to access certificate data from a TLS connection.  However, these versions of Flash were too recent to enjoy the reported 98.9\% desktop market penetration of Flash 9.0 \cite{adobe}.
Thus we implemented our tool in ActionScript using only libraries supported by the Flash 9.0 runtime.
Using the \texttt{Socket} API provided by Flash 9.0 we implemented functionality required to perform a partial TLS handshake.
After receiving the full \texttt{Certificate} message from the desired host the handshake is aborted and the connection is closed.
The Flash application records and parses all certificates received from the \texttt{Certificate} message (as some hosts offer certificate chains) and stores them locally until it parses the final one.
All certificate data, in PEM format, is concatenated and then sent as an HTTP POST request back to the host for analysis.

Code for the measurement tool and collected datasets will be available for download at \url{https://tlsresearch.byu.edu}.

\subsection{Limitations}
\label{sec:limitations}

Our tool is unable to measure TLS proxies being used against most
mobile devices.  An overwhelming majority of mobile platforms do not
support Flash and Adobe has discontinued their development of Flash
for mobile devices.

It is possible that TLS proxies could be engineered to circumvent our measurements.
At the time of our study, our measurement methodology was not well known,
so it is unlikely that any attacker was evading detection or
tampering with our reports.
However, in the case that this methodology becomes well-known, it would be difficult to prevent dedicated attackers from modifying their TLS proxies to avoid our measurements. 

\section{Google Adwords Campaigns}
\label{sec:deployment}

To achieve rapid and widespread deployment of our measurement tool we leveraged the Google AdWords platform.
This strategy for using an advertising campaign to conduct an end-user measurement study has previously been used to study CSRF attacks \cite{barth2008robust}, DNS rebinding attacks \cite{jackson2009protecting}, and DNSSEC deployment \cite{huston1,huston2,lian2013measuring}. Our study is the first to use this same method to measure the deployment of TLS proxies.
The results from this study shed light on the legitimate demand for TLS proxies as well as several suspicious or duplicitous practices.

\begin{figure}
\centering
\includegraphics[width=\columnwidth]{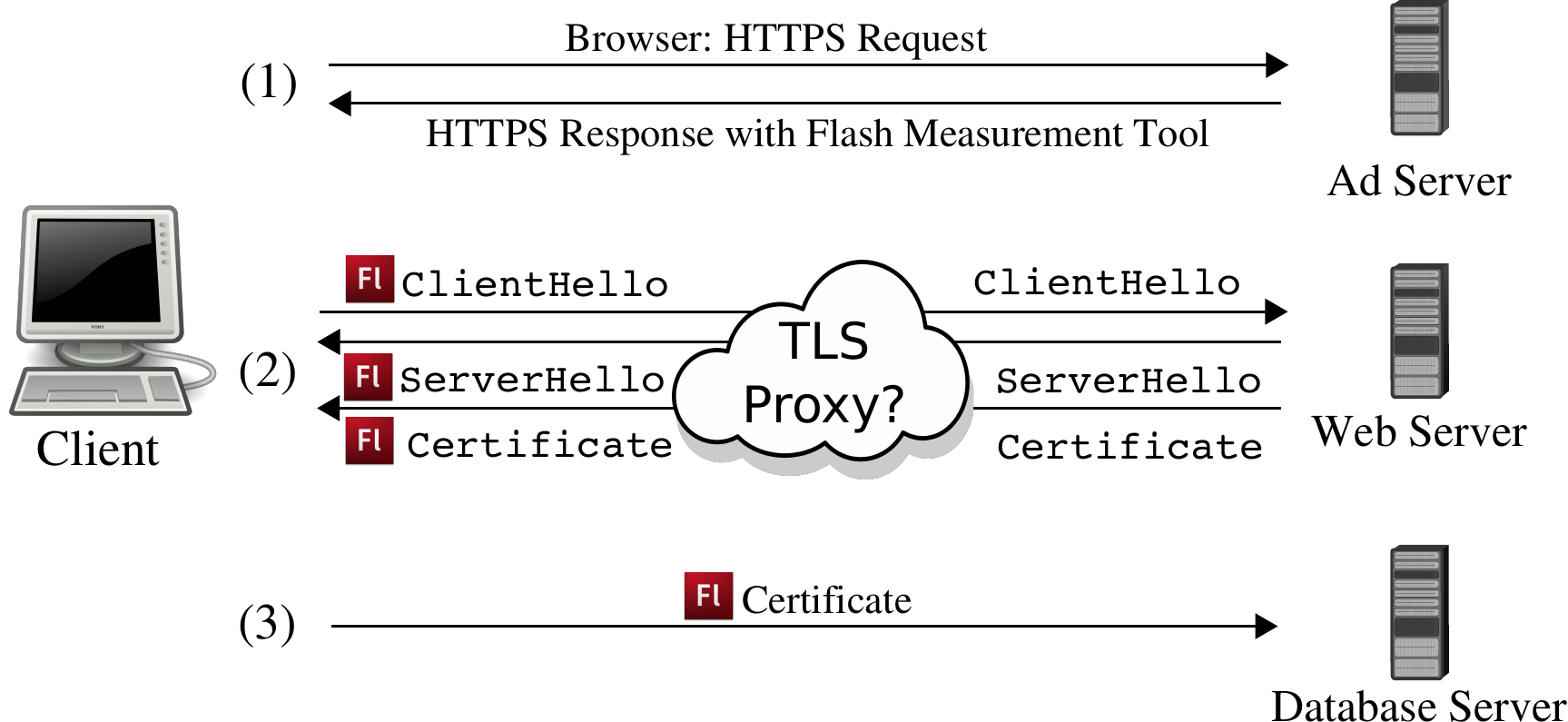}
\caption{Using an Ad Server for TLS Proxy Measurement}
\label{fig:flashtool_alt_3}
\end{figure}

\begin{figure}
\centering
\includegraphics[width=100px,height=100px]{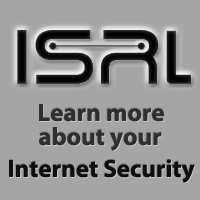}
\caption{Appearance of tool via Google AdWords}
\label{fig:adimage}
\end{figure}

The deployment of the measurement tool is given in Figure~\ref{fig:flashtool_alt_3}.
Deployment responsibilities were delegated to the Google AdWords platform, while all reports from the tool were sent back to a reporting server we controlled.
To accommodate placement in advertisements, our measurement tool was modified to contain a visible canvas on which we place a simplistic advertisement for our research lab.
Figure~\ref{fig:adimage} shows the advertisement as it appeared to web users during our measurement study.
Our measurement tool was run as soon as user's browsers loaded the advertisement, and required no interaction from the users.


For our ad campaigns we leveraged the CPM (cost-per-impression) bidding model for our campaign, which maximizes the number of unique clients presented with our ad.
We set the Max. CPM to \$10 USD.
To help us reach a global audience we indicated that our ad should be served to all locations and languages.
Additionally, since ads are shown only on websites that match a set of designated keywords we selected our keywords based on phrases that were currently trending globally on Google Trends~\cite{googletrends}.
We set our ad to show uniformly throughout the day so as to collect data from users in a variety of locations and situations (e.g., home, commuting, work).

Along with the certificate, we also recorded the IP address of the client tested.
This IP address was then used to query the MaxMind GeoLite \cite{maxmind} database to gather geolocation information.

\subsection{Campaign Setup for First Study}
Our initial Google AdWords advertising campaign ran from January 6, 2014 to January 30, 2014.
During the first 17 days of the study we varied the amount of money allocated to the ad campaign, but for the last week we kept it at \$500/day.
In this study we only gathered certificate data for our own website, \textit{tlsresearch.byu.edu}.
We used the following keywords for the study:
``Nelson Mandela'',
``Sports'',
``Basketball'',
``NSA'',
``Internet'',
``Freedom'',
``Paul Walker'',
``Security'',
``LeBron James'',
``Haiyan'',
``Snowden'',
``PlayStation 4'',
``Miley Cyrus'',
``Xbox One'',
``iPhone 5s''.

This initial campaign generated 4,634,386 impressions and 3,897 clicks (not required to complete the measurement) at a cost of \$4,911.97.
In total we completed 2,861,244 successful measurements.

Section~\ref{sec:measurementstudy} discusses the results from this first study.

\subsection{Campaign Setup for Second Study}
To increase the number of measurements collected and to better understand the nature of TLS proxies we conducted a second set of measurements approximately eight months after the first study.

One question unanswered by the first study was whether TLS proxies were intercepting all traffic, or whether they selectively intercepted traffic according to white or blacklists.
To shed light on this subject, we decided to gather measurements for different types of sites:

\begin{itemize}
\item {\bf Popular:} Sites from the Alexa top 25,000. Six websites were included in this category.
\item {\bf Business:} Commercial sites unlikely to be blocked by places of business. Five websites were included in this category.
\item {\bf Pornographic:} Pornographic websites (expected to be blocked by parental filters and places of business). Five websites are included in this category.
\item {\bf Authors':} The single website operated by the authors and also used in the first measurement study.
\end{itemize}

The policy restrictions of the Flash runtime prohibit establishment of socket connections to arbitrary hosts.
Thus all sites used in our study had to host permissive ``socket policy files'' that allowed connections to port 443 from any domain.
We scanned for the presence of permissive socket policy files on the entirety of the Alexa top 1 million websites, and selected the highest ranked such websites for each type to use in the second measurement.  Table ~\ref{tab:hosts} lists the additional hosts we probed.

\begin{table}[t]
\scriptsize
\centering
{\setlength{\extrarowheight}{2pt}
\begin{tabular}{lll}
	\toprule
	\multicolumn{3}{c}{\normalsize Website}\\
        {\normalsize Top 25,000} & {\normalsize Business} & {\normalsize Porn}\\
	\midrule
        qq.com & airdroid.com & pornclipstv.com\\
        promodj.com & webhost1.ru & porno-be.com\\
        idwebgame.com & restaurantesecia.com.br & pornbasetube.com\\
        parsnews.com & speedtest.net.in & pornozip.net\\
        idgameland.com & iprank.ir & pornorasskazov.net\\
        vcp.ir\\
        \bottomrule

\end{tabular}}
\caption{Second Study Websites Probed}
\label{tab:hosts}
\end{table}

At most 17 of these sites were queried by a single served instance of our Flash measurement tool.
Due to differences in Internet connectivity quality and hardware and software performance, not all clients served with our ad were able to successfully perform TLS handshakes with all hosts.
The tool was configured to first test the connection to the authors' website, before attempting to test connections to the other hosts in parallel.

In this second study, we also targeted specific countries by
creating an additional ad campaign for each country.
The ad image used in both the global and country specific ad campaigns was the same.
Some of the countries we wanted to target were unavailable in Google AdWords (e.g., Iran, Syria) and after discussion we settled on the following five countries: China, Egypt, Pakistan, Russia, and Ukraine.


The second study ran from October 8, 2014 4:00 PM MDT to October 15, 2014 4:00 PM MDT.
The budget for the global campaign was \$500/day and the country-specific campaigns were \$50/day.
We used the following keywords for the study:
``Nelson Mandela'',
``Sports'',
``Internet Security'',
``Basketball'',
``Football'',
``Freedom'',
``NCAA'',
``Paul Walker'',
``Boston Marathon'',
``Election'',
``North Korea'',
``Harlem Shake'',
``PlayStation 4'',
``Royal Baby'',
``Cory Monteith'',
``iPhone 6'',
``iPhone 5s'',
``Samsung Galaxy S4'',
``iPhone 6 Plus'',
``TLS Proxies''.

The breakdown of costs and results are given in Table~\ref{tab:campaign2}.
In total we completed 12,314,756 successful measurements.

\begin{table}[t]
\centering
{\setlength{\extrarowheight}{2pt}
\begin{tabular}{lccc}
	
	\toprule
	Campaign & Impressions & Click & Cost \\
	\midrule
	
	Global & 3,285,598 & 5,424 & \$4,021.78 \\
	China & 689,233 & 652 & \$401.41 \\
	Egypt & 232,218 & 1,777 & \$378.17 \\
	Pakistan & 183,849 & 2,536 & \$378.26 \\
	Russia & 230,474 & 203 & \$401.36 \\
	Ukraine & 364,868 & 294 & \$390.69 \\
	\midrule
	Total & 5,079,298 & 11,077 & \$6,090.19 \\
        \bottomrule

\end{tabular}}
\caption{Second Study Statistics}
\label{tab:campaign2}
\end{table}

Section~\ref{sec:measurementstudy2} discusses the results from this second study.

\section{Initial Measurement Study}
\label{sec:measurementstudy}

Our first measurement study was targeted at a general global audience.
During the duration of this ad campaign, we served 4.36 million ads and successfully completed 2.86 million measurements.
Of those tests, 11,764 returned a different X.509 certificate than was served by our secure web server, indicating the presence of a TLS proxy.

The users behind a proxied connection that were identified by our campaign originated in 142 countries and from 8,589 distinct IP addresses. Due to the targeting algorithms used by Google AdWords, our tool's exposure to these countries is not uniformly distributed. 
Table~\ref{tab:connectionsbycountry} shows the countries with
the most proxied connections in our study.
For each country, the table lists the total number of proxied connections, the total number of connections, and the percentage of total connections to that country that were proxied.
Some countries have significantly higher percentages of proxied connections than the average, including France (1.09\%), Canada (0.87\%), Belgium (0.81\%), the United States (0.79\%), and Romania (0.74\%).
Together, connections from the United States and Brazil account for 36\% of detected proxies.

\begin{table}[t]
\centering
{\setlength{\extrarowheight}{2pt}
\begin{tabular}{clccc}
  \toprule
	Rank & Country & Proxied & Total & Percent \\
  \midrule
	
	1 & US & 2,252 & 285,078 & 0.79\%  \\
	2 & Brazil & 2,041 & 298,618 & 0.68\% \\
	3 & France & 812 & 74,789 & 1.09\% \\
	4 & UK & 759 & 259,971 & 0.29\% \\
	5 & Romania & 696 & 94,116 & 0.74\% \\
	6 & Germany & 499 & 187,805 & 0.27\% \\
	7 & Canada & 303 & 34,695 & 0.87\% \\
	8 & Turkey & 303 & 65,195 & 0.46\% \\
	9 & India & 302 & 51,348 & 0.59\% \\
	10 & Spain & 226 & 62,569 & 0.36\% \\
	11 & Russia & 224 & 58,402 & 0.38\% \\
	12 & Italy & 200 & 129,358 & 0.15\% \\
	13 & S.Korea & 196 & 46,660 & 0.42\% \\
	14 & Portugal & 185 & 29,799 & 0.62\% \\
	15 & Poland & 182 & 110,550 & 0.16\% \\
	16 & Ukraine & 160 & 61,431 & 0.26\% \\
	17 & Belgium & 136 & 16,816 & 0.81\% \\
	18 & Japan & 111 & 31,751 & 0.35\% \\
	19 & Netherlands & 104 & 31,938 & 0.33\% \\
	20 & Taiwan & 101 & 61,195 & 0.17\% \\
	 & Other (215) & 1,972 & 869,096 & 0.23\% \\
  \bottomrule
  	 & Total & 11,764 & 2,861,180 & 0.41\% \\
  \bottomrule
\end{tabular}}

\caption{Proxied connections by country (1st study)}
\label{tab:connectionsbycountry}
\end{table}

\subsection{Analysis of Issuer Organization}

We first analyze the contents of the Issuer Organization in the
substitute certificates we collected. We use OpenSSL to
decode the certificates and store them in a database, where we can run
queries.  We also manually inspect the contents of the relevant fields
to identify the issuing organization and their software products,
using web searches to determine their identity. We emphasize that our
results in this section are based on the intercepting proxy
self-identifying themselves in the certificate. It is certainly
possible that malicious proxies have hidden their tracks by
masquerading as a legitimate organization in the Issuer Organization field, and we
cannot detect this.

Table~\ref{tab:topissuernames} shows the values for the Issuer Organization field of the substitute certificates. 
Table~\ref{tab:proxytypes} provides a breakdown of values present in the Issuer Organization field of the substitute certificates.
The majority of certificates from proxied connections have an Issuer Organization field matching the name of a personal or enterprise firewall (69.54\%).
Another 12.66\% have the name of an organization set as the Issuer Organization (e.g., Lawrence Livermore National Laboratory, Lincoln Financial Group).
Additionally, 7\% (829) of the substitute certificates have null values for the Issuer Organization field.

\begin{table}[t]
\centering
{\setlength{\extrarowheight}{2pt}
\begin{tabular}{llc}
	
	\toprule
	Rank & Issuer Organization & Connections \\
	
	\midrule

	1 & Bitdefender & 4,788 \\
	2 & PSafe Tecnologia S.A. & 1,200 \\
	3 & Sendori Inc & 966 \\
	4 & ESET spol. s r. o. & 927 \\
	5 & Null & 829 \\
	6 & Kaspersky Lab ZAO & 589 \\
	7 & Fortinet & 310 \\
	8 & Kurupira.NET & 267 \\
	9 & POSCO & 167 \\
	10 & Qustodio & 109 \\
	11 & WebMakerPlus Ltd & 95 \\
	12 & Southern Company Services & 62 \\
	13 & NordNet & 61 \\
	14 & Target Corporation & 52 \\
	15 & DigiCert Inc & 49 \\
	16 & ContentWatch, Inc. & 42 \\
	17 & NetSpark, Inc. & 42 \\
	18 & Sweesh LTD & 39 \\
	19 & IBRD & 26 \\
	20 & Cloud Services & 23 \\
	& Other (332) & 1,121 \\
	\bottomrule

\end{tabular}}
\caption{Issuer Organization field values}
\label{tab:topissuernames}
\end{table}

\begin{table}[t]
\centering
{\setlength{\extrarowheight}{2pt}
\begin{tabular}{lcc}
	
	\toprule
	Proxy Type & Connections & Percent \\
	
	\midrule

	Business/Personal Firewall & 8,101 & 68.86\% \\
	Business Firewall & 69 & 0.59\% \\
	Personal Firewall & 11 & 0.09\% \\
	Parental Control & 156 & 1.33\% \\
	Organization & 1,394 & 12.66\%\\
	School	& 32 & 0.27\% \\
	Malware	& 1,112 & 8.65\% \\
	Unknown	& 840 & 7.14\% \\
	Telecom	& 0 & 0\% \\
	Certificate Authority & 49 & 0.42\% \\
        \bottomrule

\end{tabular}}
\caption{Classification of claimed issuer in 1st study}
\label{tab:proxytypes}
\end{table}

The most suspicious activities discovered were revealed by certificates with an Issuer Organization that matched the names of malware. ``Sendori, Inc'', ``WebMakerPlus Ltd'', and ``IopFailZeroAccessCreate'' appeared in 966, 95, and 21 Issuer Organization fields, respectively.
Sendori poses as a legitimate enterprise, however they produce software that compromises the DNS lookup of infected machines, allowing them to redirect users to improper hosts. 
A TLS proxy component is used to bypass host authenticity warnings in the browser.
The substitute certificates generated by the TLS proxy are signed by a root authority that was added to the root store of the local machine at the time of infection. Substitute certificates issued by Sendori originated from 30 distinct countries.

The WebMakerPlus malware is primarily associated with inserting advertisements into Web pages.
Since modern browsers issue warnings when insecure content is queried from secure connections, we hypothesize that WebMakerPlus uses a TLS proxy to simulate that their advertisements are served from a secure connection. Substitute certificates containing markings for WebMakerPlus originated from 16 distinct countries.

Manual Internet queries revealed that malware was responsible for an Issuer Common Name field value of ``IopFailZeroAccessCreate''. The certificates containing this value originated from 14 distinct countries. Disturbingly, each certificate contained the same 512-bit public key. This malware was also reported by \cite{huang2014analyzing}.

It is somewhat surprising that these malware programs self-identify in the substitute certificates they generate, as an attacker can arbitrarily select values for the fields in a substitute certificate.

In addition to malware discoveries, we found that the names of two companies highly associated with spam were also present in numerous Issuer Organization fields.
The names ``Sweesh LTD'', and ``AtomPark Software Inc'' were found in 39 and 20 substitute certificates, respectively.
AtomPark offers tools for spammers including ``email extractors'' and ``bulk mailers''.
Sweesh offers services to spammers to overcome ``hurdles'' faced by advertisers and publishers.
Internet searches reveal that Sweesh may be responsible for the development of WebMakerPlus.

\subsection{Negligent Behavior}

Where possible, we installed and characterized personal firewall software from many of the most common companies whose names were provided in the Issuer Organization, Issuer Organizational Unit, and Issuer Common Name fields of our collected certificates.
We characterized the behavior of these solutions when running behind our own TLS proxy which issued certificates signed by an untrusted CA. While most solutions properly rejected our forged certificates, Kurupira, a parental filter that is responsible for 267 proxied connections in our dataset, did not. When visiting \texttt{google.com} and \texttt{gmail.com}, Kurupira replaced our untrusted certificate with a signed trusted one, thus allowing attackers to perform a transparent man-in-the-middle attack against Kurupira users without having to compromise root stores. In contrast, BitDefender not only blocked this forged certificate,
but also blocked a forged certificate that resolved to a new root we installed in the victim machine's root store.

Among the negligent behavior we found are TLS proxies that generate substitute certificates with weak cryptographic strength. Our original certificate has a public key size of 2048 bits. However, we find that 5,951 (50.59\%) substitute certificates have public key sizes of 1024 bits and 21 certificates have public key sizes of 512 bits.
23 (0.20\%) TLS proxies generated substitute certificates that used MD5 for signing, 21 (0.18\%) which were also 512 bit keys.
Interestingly, some TLS proxies generated certificates that have better cryptographic strength than our certificate.
Seven (0.06\%) used certificates with a key size of 2432 and five (0.04\%) used SHA-256 for signing.

In addition to problems with cryptographic strength, we discovered that 49 (0.42\%) substitute certificates claim to be signed by DigiCert, though none of them actually are. The original certificate from our secure web server is issued by DigiCert High Assurance CA-3, indicating the TLS proxy likely copied this field when creating the substitute.
It is alarming that a TLS proxy would opt to copy this field, as it signifies a masquerading as the legitimate authority.
It is possible that these proxies are operated by malicious individuals doing their best to not be detected by the user.

Finally, we note that 110 substitute certificates have modifications to the subject field. For 51 (0.43\%) certificates, the subject did not match our website's domain.
In many cases a wildcarded IP address was used that only designated the subnet of our website.
In two cases the substitute certificate is issued to the wrong domain entirely: \texttt{mail.google.com} and \texttt{urs.microsoft.com}. These
certificates appear to be legitimate for those domains and properly
validate back to GeoTrust and Cybertrust roots, respectively.

\section{Second Measurement Study}
\label{sec:measurementstudy2}

During the duration of this second ad campaign we successfully completed 12.3 million measurements targeting five specific countries as well as the world in general.
Of those tests, 50,761 returned a different X.509 certificate than was served by the authoritative host.

\subsection{Analysis of Issuer Organization}

\begin{table}[t]
\centering
{\setlength{\extrarowheight}{2pt}
\begin{tabular}{lcc}
	
	\toprule
	Proxy Type & Connections & Percent \\
	\midrule

	Business/Personal Firewall & 36,005 & 70.93\% \\
	Business Firewall & 1,231 & 2.43\% \\
	Personal Firewall & 536 & 1.06\% \\
	Parental Control & 428 & 0.84\% \\
	Organization & 3,531 & 6.96\%\\
	School	& 482 & 0.95\% \\
	Malware	& 2,571 & 5.06\% \\
	Unknown	& 5,455 & 10.75\% \\
	Telecom	& 447 & 0.88\% \\
	Certificate Authority & 68 & 0.13\% \\
        \bottomrule

\end{tabular}}
\caption{Classification of claimed issuer in 2nd study}
\label{tab:proxytypesv2}
\end{table}

Table~\ref{tab:proxytypesv2} contains the breakdown of Issuer Organization fields from our second measurement study.
As in our first measurement study, we find that the majority of TLS proxies claim to be features of firewall solutions (74.42\%).
Organization and school names are also prevalent, accounting for another 6.01\% of Issuer Organization values.
However, we see an increase in the relative popularity of the ``Unknown'' as compared to the first study (10.75\% from 7.14\%).
The Unknown category comprises certificates with null or blank issuer fields, or otherwise uncategorizable values.
In tandem with this finding we note that the Malware category has decreased in relative popularity from 8.65\% to 5.06\%.
These results were obtained largely from our five targeted countries.
The increase in the Unknown category of TLS proxies proxying these countries is alarming.
It is possible that malware using TLS proxy features in these regions is more opaque than its earlier counterparts, opting not to disclose its identity through Issuer Organization fields.
Even if this is not the case, it is alarming to note that TLS proxies may be decreasing their already-poor visibility to users in those countries.

Another distinguishing feature of our second study's Issuer Organization fields is the presence of the names of Telecom companies in the dataset.
We found 375 proxied connections from IP addresses owned by a Korean telecom company, LG UPLUS.
Another four telecom company names were reported from an additional 72 connections.
It is unclear whether LG UPLUS and the other companies are using TLS proxies within their own office buildings or using them to intercept the traffic of their own users.
The latter behavior is not without precedent; Nokia has recently come under fire for such an operation~\cite{nokia}.

\subsection{Proxy prevalence by specific country}

\begin{table}[t]
\centering
{\setlength{\extrarowheight}{2pt}
\begin{tabular}{clccc}
  \toprule
	Rank & Country & Proxied & Total & Percent \\
  \midrule
	1 & China & 563 & 2,549,301 & 0.02\%  \\
	2 & Ukraine & 4,329 & 1,575,053 & 0.27\%  \\
	3 & Russia & 4,532 & 1,116,341 & 0.40\%  \\
	4 & Korea & 1,722 & 836,556 & 0.21\%  \\
	5 & Egypt & 3,720 & 660,937 & 0.56\%  \\
	6 & Pakistan & 1,890 & 456,792 & 0.41\%  \\
	7 & Turkey & 1,975 & 411,962 & 0.48\%  \\
	8 & US & 3,327 & 385,811 & 0.86\%  \\
	9 & Japan & 2,033 & 273,532 & 0.74\%  \\
	10 & UK & 2,056 & 266,873 & 0.77\%  \\
	11 & Brazil & 1,889 & 232,454 & 0.81\%  \\
	12 & Taiwan & 530 & 186,942 & 0.28\%  \\
	13 & Romania & 2,210 & 185,749 & 1.19\%  \\
	14 & Indonesia & 798 & 181,971 & 0.44\%  \\
	15 & Germany & 1,091 & 177,586 & 0.61\%  \\
	16 & Italy & 737 & 145,438 & 0.50\%  \\
	17 & Greece & 516 & 130,613 & 0.40\%  \\
	18 & Poland & 456 & 127,806 & 0.36\%  \\
	19 & Czech Rep. & 343 & 110,170 & 0.31\%  \\
	20 & India & 716 & 102,869 & 0.70\%  \\
	 & Other (209) & 15,328 & 2,200,000 & 0.70\% \\
  \bottomrule
  	 & Total & 50,761 & 12,314,756 & 0.41\% \\
  \bottomrule
\end{tabular}}

\caption{Connections tested by country (2nd study)}
\label{tab:connectionsbycountryv2}
\end{table}

Our second measurement study via AdWords contained six mini-campaigns.
Five of these targeted the countries of China, Ukraine, Russia, Egypt, and Pakistan.
These countries were selected for their contemporary civil struggles and their respective governments' stance on free speech.
The final mini-campaign targeted the world in general.
Table~\ref{tab:connectionsbycountryv2} shows a breakdown of the number of connections tested per country.
We immediately see that all five specific countries targeted by our campaign lie within the top six most-prevalent countries, showing the dependability of Google Adwords' country targeting feature.
We also note the relatively low percentage of TLS-proxied traffic from China.
Before this study the authors hypothesized that this country would have a high amount of TLS-proxied traffic due to its stance on civil liberties and government surveillance.
We also find that Ukraine, Russia, Egypt and Pakistan all have a lower TLS-proxy percentage than western nations such as the United States (0.86\%) and the UK (0.77\%).
This may be due to the fact that most detected firewall solutions are of western origin and western install base.
Thus it is possible that the general lower proxy rates in our target countries is due more to consumer choice and buying power.

\begin{figure}
\centering
\includegraphics[width=240px]{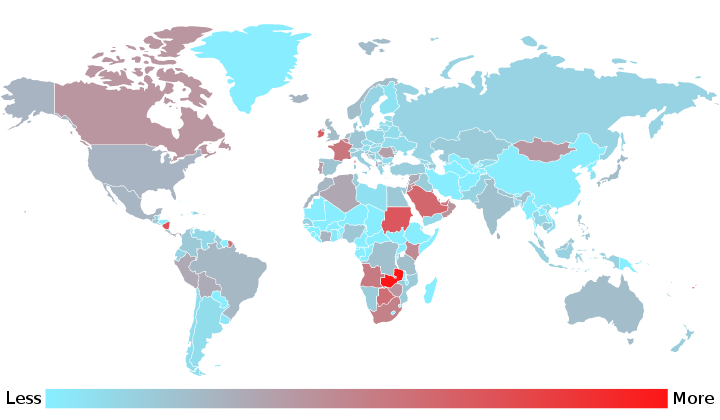}
\caption{Heat-map of TLS proxy prevalence by country. Highest = 12\% proxy rate, lowest = 0\% proxy rate}
\label{fig:worldheatmap}
\end{figure}

The relative prevalence of TLS proxies by country is visualized in Figure~\ref{fig:worldheatmap}.
Low TLS-proxy rates are signified by blue and gradually transition to red with increasing proxy rate.
The map includes connection data from our second study in 228 countries and territories.

\subsection{Proxy behavior by type of host}

\begin{table}[t]
\centering
{\setlength{\extrarowheight}{2pt}
\begin{tabular}{lccc}
	
	\toprule
	Website Type & Connections & Proxied & Percent Proxied \\
	
	\midrule

Popular & 5,132,342 & 20,965 & 0.41\% \\
Business & 1,787,875 & 7,494 & 0.42\% \\
Pornographic & 3,004,996 & 12,458 & 0.41\% \\
Authors' & 2,353,717 & 9,844 & 0.42\% \\

        \bottomrule

\end{tabular}}
\caption{Proxied connection breakdown by host type}
\label{tab:proxied_by_host_type}
\end{table}

The augmented measurement tool used in our second study connected to various types of hosts: popular, business, pornographic, and our own.
The breakdown of the prevalence of TLS proxies with respect to each host type is shown in Table~\ref{tab:proxied_by_host_type} (note that the number of connections to each type of host is varies due to the variance in bandwidth and computer performance of our users). 
The percentage of proxied traffic to each type of host is nearly identical.
We also find that individual TLS proxies are also indiscriminate in their behavior toward these types of hosts.
These results suggest that TLS proxies do not employ blacklists when deciding which traffic to intercept.
Given Huang et al.'s finding of a 0.20\% TLS-proxy rate for Facebook connections, there is some evidence to suggest that at least some TLS proxies are employing whitelists when determining whether to intercept a connection.
Facebook's popularity far exceeds the popularity of our chosen popular hosts (we were constrained to hosts which served permissive Flash socket policy files), so sites akin to it are in a class of their own.
It is possible that many benevolent TLS proxies are configured to ignore extremely popular websites run by reputable organizations, perhaps to preserve some privacy and reduce performance overhead.

\subsection{More malware}

Our second study revealed a continued presence of malware in the TLS proxy space.
All of our previously discovered malware was also present in our second study, with an additional five discoveries: issuer fields containing the values Objectify Media Inc (1069 connections), Superfish, Inc. (610 connections), WiredTools LTD (131 connections), Internet Widgits Pty Ltd (67 connections), and ImpressX OU (16 connections).
Web research indicated that all these are malware products and one of them, Internet Widgits Pty Ltd, has ties to a botnet.
Combined with malware previously identified, malware in the second study accounted for 2,571 of proxied connections.

One suspicious Issuer Organization field was ``kowsar''. Certificates with this identifier appeared 268 times, and
were retrieved by 266 unique IP addresses. Unlike other Issuer
Organizations we found, this identifier did not appear to be
associated with a large organization (which would indicate a corporate
firewall) or a personal firewall product. The IP addresses given this
certificate are from numerous countries around the world and from many
different ISPs.  Contrast this with the Certificate Issuer ``DSP'',
which appeared 204 times, but with only one IP. In this case, ``DSP''
is being used by the Department of Social and Family Affairs (also
called the Department of Social Protection), Ireland, and thus likely
represents a corporate firewall. The pattern for ``kowsar'' is
indicative of either a popular personal firewall or an active attack
such as a botnet.

Similar oddities appear, but on a smaller scale. For example, the
Certificate Issuer field ``Information Technology'' appeared 33 times,
covering 3 IP addresses. These IPs were from a Japanese chemical
company, an ISP in Netherlands, and a chapter of the American Red
Cross. These are such disparate organizations that this looks like
suspicious behavior, though it is possible that each of these
organizations set up a corporate firewall and chose the same name for
the certificate they generated. The Issuer field ``MYInternetS''
appeared 36 times from 6 different ISPs. Five of these are in Denmark,
from a variety of ISPs and a university, yet one is from a cable
subscriber in the United States. It is difficult to determine whether
cases like this are examples of malware.

Even more worrisome are the 5,455 instances where we could not identify the issuer, and
the 1,518 where the issuer field is null or blank.  Whoever set up the TLS proxy in these cases did not want to be
identified.

\section{Mitigation}

A large body of work seeks to detect and prevent TLS proxies,
generally regarding them as MitM attacks.  Clark and van
Oorschot~\cite{clark2013sok} provide an extensive survey of this area
and provide one of the few research papers that acknowledges the
existence of benevolent TLS proxies. Below we survey the various
mitigation approaches in the field.

{\em Multi-path probing} allows clients to determine whether the
certificate they have been given for a server is different from those
seen by most other clients. Representative systems include
Perspectives~\cite{wendlandt2008perspectives},
Convergence~\cite{marlinspike2011ssl}, and
DoubleCheck~\cite{alicherry2009doublecheck}.
Crossbear~\cite{holz2012x} goes a step further to use traceroute to
localize the origin of the attack. Other systems use existing
Certificate Authorities or centralized notaries to vouch for the
authenticity of a certificate \cite{mecai,amann2012extracting,
  amann2012revisiting}. Each of these systems may suffer from false
alarms due to benign changes to certificates and the presence of
multiple valid certificates for a given site \cite{amann2013acsac}.

There are several proposals to leverage the {\em shared password}
between the client and server to prevent a MitM attack.  Direct
Validation of Certificates (DVCert)~\cite{dacosta2012trust} permits
the server to attest to the authenticity of all the certificates used
in a session with the web application, including certificates from
other domains.  SSL/TLS Session-Aware User
Authentication~\cite{oppliger2006ssl,oppliger2008ssl} thwarts TLS MitM
attacks through user authentication tokens based on client credentials
and TLS session information.  The proposed TLS-SRP
protocol~\cite{rfc5054} would extend the TLS handshake to support
mutual authentication based on a shared password.

Several proposals leverage {\em DNS} to prevent MitM attacks.
ConfiDNS~\cite{poole2006confidns} utilizes the temporal and spatial
redundancy of the existing DNS system to assess agreement for IP
resolution.  The DNS-Based Authentication of Named Entities
(DANE)~\cite{hoffman2011using} protocol enables administrators to bind
hostnames to their certificates.  This permits public keys to be
transmitted via DNSSEC without involving a CA.

{\em Certificate pinning}~\cite{evans2011certificate} is a Google
proposal for the web server to limit all future HTTPS connections to a
limited set of server certificates. Pinning is a trust-on-first-use
technology.  The Google Chrome browser comes pre-configured with some
Google certificates already pinned in advance so the user does not
have the TOFU issue with those sites.  Chrome also trusts any locally
installed trusted roots, so benevolent proxies and malware can
circumvent the pinning process.  Trust Assertions for Certificate Keys
(TACK)~\cite{tack} is a TLS extension for the server to pin a signing
key that must sign all other keys in the domain.

Another approach is to use an {\em audit log} of valid certificates
issued by Certificate Authorities. Representative work in this area is
Certificate Transparency (CT)~\cite{rfc6962,ryan2014enhanced} and the
EFF sovereign keys (SK) project~\cite{sovereignkeys}.  The Accountable
Key Infrastructure (AKI)~\cite{kim2013accountable} is a proposal for
new infrastructure to validate public keys and reduce the reliance on
CAs.  This system also includes public log servers that support public
validation of certificate integrity.

A different approach acknowledges that there is an industry need for
TLS inspection to detect malware or protect intellectual
property.  Several proposals to the IETF from industry introduce
mechanisms that would make proxies visible to the other participants
in a chain of TLS connections and could include user notification and
consent \cite{mcgrew2012internetdraft, loreto2014internetdraft}.
Another proposal \cite{nir2012internetdraft} introduces an extension
to TLS that enables a client to share decryption keys with a TLS
proxy. A potential drawback is that the key-sharing mechanism
represents a new attack surface that hackers could attempt to exploit
to acquire the decryption keys for a TLS session.

\section{Related Work}

The most closely related work in this field is a recent paper by Huang
et al., which independently develops a measurement tool that is very
similar to ours and conducts a measurement study of TLS proxies that
intercept the Facebook website \cite{huang2014analyzing}. We discussed
the primary differences in methodology in
Section~\ref{sec:introduction}.  Generally speaking, the advantage of
Huang's methodology is that they find proxies specifically targeting
Facebook, whereas the advantage of our methodology is that we can
target our measurements for selected countries and for selected
websites that have permissive Flash socket policy files. This enables
us to actively collect a broader measurement of proxies.

In comparing our results to Huang, the prevalence of proxies in our
study is roughly twice what was measured by Huang (0.41\% versus
0.20\%). In addition, we find a wider array of malware, deceptive
practices, and suspicious circumstances. Both of these results are
likely due to the more comprehensive measurements we make, avoiding a
site such as Facebook that is likely on the whitelist for many
proxies. Our measurements of WebMakerPlus, Objectify Media, Superfish,
WiredTools, Internet Widgits Pty, ImpressX, and kowsar all represent
malware found only in our study. Likewise, the presence of spam
infections from Sweesh and AtomPark are unique to our study, as is the
evidence of botnets using TLS proxies. We are the first to identify a
parental filter replacing an untrusted certificate with a trusted
one. Our country-specific measurement campaigns add additional data to
the field.

The only other paper to find evidence of TLS proxies is the work from
The Netalyzer project, which analyzes the root store of Android
devices \cite{vallina2014tangled}. Their primary findings include the
use of manufacturer and vendor-specific certificates, the presence of
unusual root certs, and third party apps that manipulate the root
store. In addition, they find one case of a TLS proxy, out of 15,000
assessed TLS sessions. The app whitelists several sites, including
Facebook, Twitter, and several Google sites, but intercepts mail from
Yahoo, Google, and traffic to several major banks. It is difficult
to compare the prevalence (1 in 15K) to rates found by Huang and this
paper because the sample is from users choosing to download the
Netalyzer App.

Another closely related paper is the Crossbear
system~\cite{holz2012x}, which is designed for volunteer hunters to
work together to detect and localize real-world TLS MitM attacks.
After the client establishes a TLS connection to a website, the client
sends the received certificate chain to a central Crossbear server.
The Crossbear server establishes its own secure connection with the
website and also queries Convergence for additional data about the
website's certificate.  This information is recorded in a database on
the server and is also sent to the client.  If the cumulative data
received by the client suggest a MitM is present, the client performs
a traceroute operation to the malicious server and sends that
information to the Crossbear server.  The Crossbear server attempts to
localize the origin of the MitM attacker by using traceroute data from
many Crossbear clients. Crossbear was deployed in 150 locations on the
PlanetLab testbed and had not detected any attacks (or benevolent TLS
proxies) at the time of the report.

Finally, a number of surveys collect and analyze SSL certificates and
certificate authorities on the Internet \cite
{eckersley2011decentralized,holz2011ssl,akhawe2013here,durumeric2013analysis,amann2013acsac}. These
studies do not examine the use of substitute certificates by TLS
proxies, but focus on issues such as TLS errors, properties of
certificates and the PKI system, and poor security practices.

\section {Conclusion}

Using an automatic measurement tool and two large-scale measurement campaigns via Google AdWords, we expose a variety of TLS proxies worldwide.
We find that overall, 0.41\% of all connections tested are behind a TLS proxy.
Given that both benevolent and malicious uses of TLS proxies use similar if not identical methodologies, distinguishing between the two is a difficult task.
For example, analysis of substitute certificate fields shows that most TLS proxies claim to be acting on behalf of concerned users, behaving as firewalls for both personal and business use.
However, since TLS proxies violate the normal hierarchy of trust, it is impossible to verify the claims made in those fields.
Despite this, we have found eight distinct, self-identifying malware which proxied over 3,600 of our total connections.
The prevalence of malware using TLS proxying techniques illustrates the need for stronger controls over the root stores of browsers and operating systems.
Even more TLS proxy instances chose to remain entirely anonymous by providing indiscernible or no information in substitute certificate data.
Our additional findings of telecom-run TLS proxies, null issuer fields, and falsified certificate authority signatures further highlight the need for transparency in this area.
We stress the need for better systems in the browser and/or operating system to assist in both user awareness and distinguishing between benevolent and malicious uses of TLS proxies.


\section{Acknowledgments}

This work is supported by a 2014 Google Faculty Research Award.

\bibliographystyle{abbrv}
\bibliography{bib}

\begin{thebibliography}{10}

\bibitem{iranattack}
H.~Adkins.
\newblock An update on attempted man-in-the-middle attacks.
\newblock
  \url{http://googleonlinesecurity.blogspot.com/2011/08/update-on-attempted-man-in-middle.html}.

\bibitem{adobe}
Adobe.
\newblock Adobe {Flash Player PC} penetration.
\newblock
  \url{http://www.adobe.com/products/player_census/flashplayer/PC.html}.
\newblock Accessed: 22 March, 2013.

\bibitem{akhawe2013here}
D.~Akhawe, B.~Amann, M.~Vallentin, and R.~Sommer.
\newblock Here's my cert, so trust me, maybe?: understanding {TLS} errors on
  the web.
\newblock In {\em Proceedings of the 22nd international conference on World
  Wide Web}, pages 59--70. International World Wide Web Conferences Steering
  Committee, 2013.

\bibitem{alicherry2009doublecheck}
M.~Alicherry and A.~D. Keromytis.
\newblock Doublecheck: Multi-path verification against man-in-the-middle
  attacks.
\newblock In {\em 14th IEEE Symposium on Computers and Communications (ISCC)},
  pages 557--563. IEEE, 2009.

\bibitem{amann2013acsac}
B.~Amann, R.~Sommer, M.~Vallentin, and S.~Hall.
\newblock No attack necessary: The surprising dynamics of {SSL} trust
  relationships.
\newblock In {\em Proceedings of the 29th Annual Computer Security Applications
  Conference}, ACSAC '13, pages 179--188, New York, NY, USA, 2013. ACM.

\bibitem{amann2012extracting}
B.~Amann, M.~Vallentin, S.~Hall, and R.~Sommer.
\newblock Extracting certificates from live traffic: A near real-time {SSL}
  notary service.
\newblock Technical report, TR-12-014, ICSI Nov. 2012, 2012.

\bibitem{amann2012revisiting}
B.~Amann, M.~Vallentin, S.~Hall, and R.~Sommer.
\newblock Revisiting {SSL}: A large-scale study of the internet's most trusted
  protocol.
\newblock Technical report, TR-12-015, ICSI Dec. 2012, 2012.

\bibitem{barth2008robust}
A.~Barth, C.~Jackson, and J.~C. Mitchell.
\newblock Robust defenses for cross-site request forgery.
\newblock In {\em Proceedings of the 15th ACM Conference on Computer and
  Communications Security}, pages 75--88. ACM, 2008.

\bibitem{bluecoat}
T.~Chiu.
\newblock The growing need for {SSL} inspection.
\newblock
  \url{http://www.bluecoat.com/security/security-archive/2012-06-18/growing-need-ssl-inspection/},
  2011.
\newblock Accessed: 27 February , 2014.

\bibitem{clark2013sok}
J.~Clark and P.~C. van Oorschot.
\newblock {SoK}: {SSL} and {HTTPS}: Revisiting past challenges and evaluating
  certificate trust model enhancements.
\newblock In {\em IEEE Symposium on Security and Privacy (SP),}, pages
  511--525. IEEE, 2013.

\bibitem{bluecoatproxysg}
B.~Coat.
\newblock Proxysg.
\newblock \url{http://www.bluecoat.com/products/proxysg}.
\newblock Accessed: 9 January, 2014.

\bibitem{dacosta2012trust}
I.~Dacosta, M.~Ahamad, and P.~Traynor.
\newblock Trust no one else: Detecting {MITM} attacks against {SSL/TLS} without
  third-parties.
\newblock In {\em Computer Security--ESORICS 2012}, pages 199--216. Springer,
  2012.

\bibitem{durumeric2013analysis}
Z.~Durumeric, J.~Kasten, M.~Bailey, and J.~A. Halderman.
\newblock Analysis of the {HTTPS} certificate ecosystem.
\newblock In {\em Internet Measurement Conference}, 2013.

\bibitem{syriaattack}
P.~Eckersley.
\newblock A syrian man-in-the-middle attack against facebook.
\newblock
  \url{https://www.eff.org/deeplinks/2011/05/syrian-man-middle-against-facebook},
  May 2011.

\bibitem{eckersley2011decentralized}
P.~Eckersley and J.~Burns.
\newblock The (decentralized) {SSL} observatory.
\newblock In {\em USENIX Security Symposium}, 2011.

\bibitem{sovereignkeys}
E.~F.~F. (EFF).
\newblock {The Sovereign Keys Project}.
\newblock \url{http:/www.eff.org/sovereign-keys/}, 2011.

\bibitem{mecai}
K.~Engert.
\newblock {MECAI} - mutually endorsing {CA} infrastructure.
\newblock \url{http://kuix.de/mecai}.
\newblock Accessed: March 2013.

\bibitem{evans2011certificate}
C.~Evans and C.~Palmer.
\newblock Certificate pinning extension for {HSTS}.
\newblock \url{http://tools.ietf.org/html/draft-evans-palmer-hsts-pinning-00}.
\newblock Accessed: 22 March, 2013.

\bibitem{googletrends}
Google.
\newblock Google trends globally trending keywords.
\newblock \url{http://www.google.com/trends/?geo}.
\newblock Accessed: 01 January, 2014.

\bibitem{adlocksmith}
A.~Group.
\newblock {AD} locksmith.
\newblock \url{http://www.accessdata.com/products/cyber-security/ad-locksmith}.
\newblock Accessed: 9 January, 2014.

\bibitem{hoffman2011using}
P.~Hoffman and J.~Schlyter.
\newblock The {DNS}-based authentication of named entities ({DANE}) transport
  layer security {(TLS)} protocol: {TLSA, RFC 6698}.
\newblock \url{https://datatracker.ietf.org/doc/rfc6698}, 2012.
\newblock Accessed: 24 Feb, 2014.

\bibitem{holz2011ssl}
R.~Holz, L.~Braun, N.~Kammenhuber, and G.~Carle.
\newblock The {SSL} landscape: a thorough analysis of the x.509 {PKI} using
  active and passive measurements.
\newblock In {\em Proceedings of the 2011 ACM SIGCOMM conference on Internet
  measurement conference}, pages 427--444. ACM, 2011.

\bibitem{holz2012x}
R.~Holz, T.~Riedmaier, N.~Kammenhuber, and G.~Carle.
\newblock X.509 forensics: Detecting and localising the {SSL/TLS}
  men-in-the-middle.
\newblock In {\em 17th European Symposium on Research in Computer Security
  (ESORICS)}, pages 217--234. Springer, 2012.

\bibitem{huang2014analyzing}
L.-S. Huang, A.~Rice, E.~Ellingsen, and C.~Jackson.
\newblock Analyzing forged ssl certificates in the wild.
\newblock In {\em To appear, IEEE Symposium on Security and Privacy}, 2014.

\bibitem{huston1}
G.~Huston.
\newblock Counting {DNSSEC}.
\newblock \url{https://labs.ripe.net/Members/gih/counting-dnssec}.
\newblock Accessed: 26 February, 2014.

\bibitem{huston2}
G.~Huston and G.~Michaelson.
\newblock Measuring {DNSSEC} performance.
\newblock \url{http://potaroo.net/ispcol/2013-05/dnssec-performance.html}.
\newblock Accessed: 26 February, 2014.

\bibitem{jackson2009protecting}
C.~Jackson, A.~Barth, A.~Bortz, W.~Shao, and D.~Boneh.
\newblock Protecting browsers from {DNS} rebinding attacks.
\newblock {\em ACM Transactions on the Web (TWEB)}, 3(1):2, 2009.

\bibitem{kim2013accountable}
T.~H.-J. Kim, L.-S. Huang, A.~Perring, C.~Jackson, and V.~Gligor.
\newblock Accountable key infrastructure {(AKI)}: A proposal for a public-key
  validation infrastructure.
\newblock In {\em Proceedings of the 22nd International Conference on World
  Wide Web}, pages 679--690. International World Wide Web Conferences Steering
  Committee, 2013.

\bibitem{rfc6962}
B.~Laurie, A.~Langley, and E.~Kasper.
\newblock Certificate transparency, {IETF RFC 6962}.
\newblock \url{http://tools.ietf.org/html/rfc6962}, Jun 2013.

\bibitem{lian2013measuring}
W.~Lian, E.~Rescorla, H.~Shacham, and S.~Savage.
\newblock Measuring the practical impact of {DNSSEC} deployment.
\newblock In {\em Proceedings of USENIX Security}, 2013.

\bibitem{loreto2014internetdraft}
S.~Loreto, J.~Mattsson, R.~Skog, H.~Spaak, G.~Gus, and M.~Hafeez.
\newblock Explicit trusted proxy in {HTTP/2.0, Internet Draft}.
\newblock
  \url{http://tools.ietf.org/html/draft-loreto-httpbis-trusted-proxy20-01},
  February 2014.

\bibitem{marlinspike2011ssl}
M.~Marlinspike.
\newblock {SSL} and the future of authenticity.
\newblock {\em Black Hat USA}, 2011.

\bibitem{tack}
M.~Marlinspike and T.~Perrin.
\newblock Trust assertions for certificate keys.
\newblock \url{http://tack.io/}, 2013.

\bibitem{maxmind}
MaxMind.
\newblock Geolite.
\newblock \url{http://dev.maxmind.com/geoip/legacy/geolite/#IP_Geolocation}.
\newblock Accessed: 27 February, 2014.

\bibitem{mcgrew2012internetdraft}
D.~McGrew, D.~Wing, Y.~Nir, and P.~Gladstone.
\newblock {TLS} proxy server extension, {Internet-Draft, TLS Working Group}.
\newblock \url{http://tools.ietf.org/html/draft-mcgrew-tls-proxy-server-01},
  July 2012.

\bibitem{nokia}
D.~Meyer.
\newblock Nokia: Yes, we decrypt your {HTTPS} data, but don't worry about it.
\newblock \url{http://gigaom.com/2013/01/10/nokia-yes-we-decryptyour-
  https-data-but-dont-worry-about-it/}.
\newblock Accessed: 13 April, 2013.

\bibitem{paloalto}
P.~A. Networks.
\newblock Decryption.
\newblock
  \url{https://www.paloaltonetworks.com/products/features/decryption.html}.
\newblock Accessed: 27 February, 2014.

\bibitem{nir2012internetdraft}
Y.~Nir.
\newblock A method for sharing record protocol keys with a middlebox in {TLS},
  {Internet-Draft, TLS Working Group}.
\newblock \url{http://http://tools.ietf.org/html/draft-nir-tls-keyshare-02},
  March 2012.

\bibitem{oppliger2006ssl}
R.~Oppliger, R.~Hauser, and D.~Basin.
\newblock {SSL/TLS} session-aware user authentication--or how to effectively
  thwart the man-in-the-middle.
\newblock {\em Computer Communications}, 29(12):2238--2246, 2006.

\bibitem{oppliger2008ssl}
R.~Oppliger, R.~Hauser, and P.~D. BASIN.
\newblock {SSL/TLS} session-aware user authentication.
\newblock {\em Computer}, 41(3):59--65, 2008.

\bibitem{poole2006confidns}
L.~Poole and V.~S. Pai.
\newblock {ConfiDNS}: Leveraging scale and history to improve {DNS} security.
\newblock In {\em Proceedings of WORLDS}, volume~6, 2006.

\bibitem{ryan2014enhanced}
M.~D. Ryan.
\newblock Enhanced certificate transparency and end-to-end encrypted mail.
\newblock In {\em Network and Distributed System Security Symposium (NDSS)}.
  Internet Society, 2014.

\bibitem{symantecwebgateway}
Symantec.
\newblock Web gateway.
\newblock \url{http://www.symantec.com/web-gateway}.
\newblock Accessed: 9 January, 2014.

\bibitem{rfc5054}
D.~Taylor, T.~Wu, and T.~Perrin.
\newblock Using the secure remote password {(SRP)} protocol for {TLS}
  authentication, {IETF} {RFC} 5054, {Network Working Group}.
\newblock \url{http://tools.ietf.org/html/rfc5054}, Nov 2007.

\bibitem{vallina2014tangled}
N.~Vallina-Rodriguez, J.~Amann, C.~Kreibich, N.~Weaver, and V.~Paxson.
\newblock A tangled mass: The android root certificate stores.
\newblock In {\em Proceedings of the 10th ACM International on Conference on
  emerging Networking Experiments and Technologies}, pages 141--148. ACM, 2014.

\bibitem{wendlandt2008perspectives}
D.~Wendlandt, D.~G. Andersen, and A.~Perrig.
\newblock Perspectives: Improving {SSH}-style host authentication with
  multi-path probing.
\newblock In {\em USENIX Annual Technical Conference}, pages 321--334, 2008.

\end{thebibliography}

\end{document}